\documentclass[%
 reprint,
 amsmath,amssymb,
 aps,
]{revtex4-2}
\usepackage{float}
\usepackage{graphicx}
\usepackage{dcolumn}
\usepackage{bm}
\usepackage{booktabs}
\usepackage{makecell}

\makeatletter
\def\@seccntformat#1{\csname the#1\endcsname.\quad}
\renewcommand\section{\@startsection {section}{1}{0pt}%
  {-3.5ex \@plus -1ex \@minus -.2ex}%
  {2.3ex \@plus.2ex}%
  {\normalfont\normalsize\bfseries\raggedright}}
\renewcommand\subsection{\@startsection {subsection}{2}{0pt}%
  {-3.25ex\@plus -1ex \@minus -.2ex}%
  {1.5ex \@plus .2ex}%
  {\normalfont\normalsize\bfseries\raggedright}}
\makeatother


\begin{document}


\title{Heterogeneously Integrated Efficient and Widely Tunable Lasers at 795~nm for Rubidium-Based Quantum Technologies}




\author{Max~Kiewiet$^{1,2,*}$, Stijn~Cuyvers$^{1,2}$, Tom~Reep$^{1,2}$, Konstantinos~Akritidis$^{1,2}$, Gaudhaman~Jeevanandam$^{2}$, Manuel~Reza$^{2}$, Roelof~Jansen$^{2}$, Pol~Van~Dorpe$^{2}$, G\"unther~Roelkens$^{1,2}$, Kasper~Van~Gasse$^{1,2}$ and Bart~Kuyken$^{1,2,\dagger}$\\  
\vspace{+0.1 in}
\textit{\small{
$^1$Photonics Research Group, INTEC, Ghent University - imec, 9052 Ghent, Belgium\\
$^2$imec, Kapeldreef 75, 3001 Leuven, Belgium. \\
{\small $^*$Max.Kiewiet@ugent.be, $^\dagger$Bart.Kuyken@ugent.be}}}}

\date{6 August 2026}

\begin{abstract}
Scaling quantum processors and optical atomic clocks fundamentally requires orders-of-magnitude reductions in the size, weight, power, and cost of optical control systems. Photonic integration of lasers is critical to fulfill these requirements. At the near-infrared wavelengths required for atomic clocks and quantum computing through manipulation of rubidium atoms, laser integration is hindered by difficulty in light coupling and poor heat dissipation. Here, we introduce a wafer-scalable method utilizing micro-transfer printing to integrate GaAs-based amplifiers in etched recesses, directly butt-coupled to silicon nitride waveguides. We demonstrate extended-cavity single-mode lasers using this integration approach. Our compact microgear laser achieves a narrow 3~kHz fundamental linewidth at an on-chip output power of $>$22~mW---a record for a single-mode heterogeneously integrated laser in the 780--800~nm band---with a wall-plug efficiency of 9.4~\%, showcasing the high-power and efficiency potential of this integration approach. Additionally, we demonstrate a widely tunable laser leveraging Vernier filters to achieve lasing with 9~nm coarse tuning, a quasi-continuous fine-tuning range exceeding 140~GHz, and a mode-hop-free tuning range of 45~GHz. Our scalable integrated laser toolkit shows great promise for replacing macroscopic external-cavity diode lasers in next-generation quantum technologies and optical atomic clocks.

\end{abstract}
\maketitle

\mbox{}
\clearpage
\newpage

\onecolumngrid
\begin{center}
    \includegraphics[]{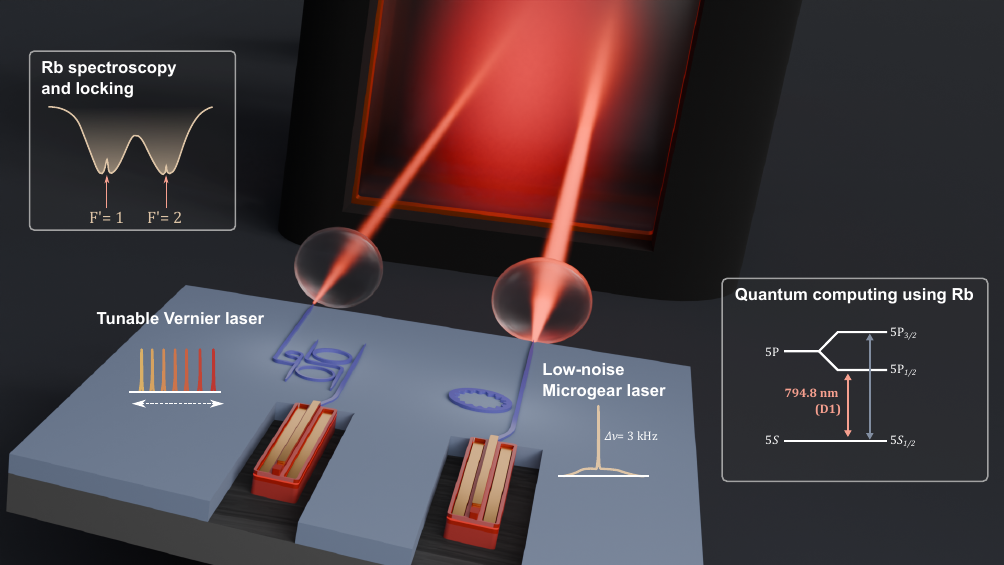}
    \makeatletter
    \def\@captype{figure}
    \makeatother
    \caption{\textbf{Scaling Rubidium-Based Quantum Control Systems.} An outlook schematic of the tunable Vernier laser and the low-noise microgear laser shining at a Rb vapor cell with highlighted applications (insets): Rb spectroscopy and wavelength locking; interaction with the Rb D1 line for quantum computing. The D1 wavelength labeled 794.8~nm in the figure is the air wavelength, corresponding to 795.0~nm in vacuum.}
    \label{fig:introduction}
\end{center}
\twocolumngrid

\section{Introduction}

 Miniaturizing complex optical control systems for neutral atom and ion interaction via photonic integrated circuits (PICs) is recognized as a key hardware bottleneck for scaling quantum processors beyond thousands of physical qubits \cite{lu2024emerging}. Neutral atom interaction with rubidium (Rb) is at the forefront of modern quantum technologies, driving critical advancements in fault-tolerant quantum computing \cite{browaeys2020many} and highly deployable optical atomic clocks \cite{maurice2020miniaturized, newman2019architecture}. Currently, macroscopic external-cavity diode lasers (ECDLs) and bulk free-space optics dominate atom manipulation. Translating these laboratory-scale demonstrations into scalable, field-ready architectures fundamentally requires orders-of-magnitude reductions in size, weight, power, and cost (SWaP-C).

The physical operations within Rb-based quantum systems impose exceptionally stringent constraints on laser sources, particularly near the D1 ($5S_{1/2} \rightarrow 5P_{1/2}$) and D2 ($5S_{1/2} \rightarrow 5P_{3/2}$) principal transitions at 795~nm and 780~nm, respectively \cite{krupke2003resonance, brown2019gray, grier2013lambda, colzi2016sub}. These processes demand continuous-wave (CW) lasers with sub-100-kHz fundamental linewidths---well below the $\sim 6$~MHz natural linewidth of the principal atomic transitions \cite{heavens1961radiative}---and on-chip output powers exceeding 10~mW to ensure rapid state preparation and efficient atomic cooling. For higher-fidelity quantum systems using e.g. Rydberg states \cite{gaetan2009observation, urban2009observation}, even more stringent requirements are placed on laser linewidth, requiring sub-kilohertz, or ideally 10-Hz-level intrinsic linewidths \cite{jiang2023sensitivity, yakshina2018three}.

Furthermore, translating these technologies to the field necessitates robust absolute frequency referencing. By resolving the narrow hyperfine structure of the D1 or D2 transitions, Doppler-free techniques yield spectral features that serve as ideal discriminators for precise laser frequency stabilization and time references \cite{debs2008piezo, strangfeld2022compact,newman2019architecture}.

To fulfill the rigorous SWaP-C and coherence requirements for solving the laser hardware bottleneck for scaling such applications, various photonic integration strategies have been actively pursued in the 780--800~nm near-infrared window. Purely monolithic integration on native III-V substrates (e.g., GaAs, InP) provides robust active devices but suffers from relatively high internal propagation losses, capping the cavity quality factor ($Q$) and thereby limiting the minimum achievable linewidth \cite{komljenovic2015widely, liang2021recent}. Hybrid integration circumvents this limitation by edge-coupling separate III-V gain chips to ultra-low-loss dielectric platforms like silicon nitride (SiN) to create extended-cavity lasers, achieving sub-Hz fundamental laser linewidths via self-injection locking \cite{isichenko2024sub}, but at the cost of highly complex and difficult-to-scale active chip alignment.

Recently, fully heterogeneous integration has emerged as a highly scalable and manufacturable alternative. Building on earlier heterogeneous frameworks, researchers have successfully demonstrated state-of-the-art fully integrated GaAs-on-SiN lasers in the sub-micrometer \cite{tran2022extending} and 780--800~nm band \cite{zhang2023photonic, castro2025integrated,thiel2026wafer} via direct wafer bonding, achieving $<$6~kHz intrinsic linewidths, wide mode-hop-free tuning ranges exceeding 100~GHz, and the co-integration of modulators and photodetectors. However, traditional wafer bonding at these shorter wavelengths requires complex evanescent coupling tapers which require high fabrication tolerances, especially at shorter wavelengths where critical dimensions decrease. Furthermore, wafer-bonded lasers inherently suffer from poor thermal dissipation through the thick oxide bottom cladding underneath the integrated III-V material. This thermal isolation, combined with sub-optimal epitaxial layer design requirements imposed to facilitate evanescent coupling, results in reduced wall-plug efficiencies (WPE) and a lower maximum output power due to thermal rollover. This means that wafer-bonded heterogeneous lasers, while well-suited for most lower-power applications, are limited in their application potential where higher output powers are required. 

Using the more flexible micro-transfer printing (MTP) approach \cite{roelkens2024present, chen2026micro}, fully pre-fabricated GaAs-based reflective semiconductor optical amplifiers (RSOAs) can be butt-coupled directly into etched recesses on a SiN platform, enabling highly efficient, wavelength-agnostic optical coupling and superior thermal dissipation directly to the silicon substrate \cite{justice2012wafer, loi2018thermal}. First demonstrations of this technique showed butt-coupling of stand-alone Fabry-Perot laser coupons to silicon photonics in the C- and O-bands \cite{uzun2023integration, juvert2018integration} reaching up to single mW-level output powers on-chip with $<$10~\% coupling efficiencies to the silicon photonics platform. Recently, the same approach was used to reach $>$100~mW power divided over two SiN waveguides \cite{hu2026greater}, showing the high-power and efficiency potential of this technique, leveraging the superior heat-sinking and more optimal epitaxial layer design compared to evanescently coupled laser integration. To show the potential for more complex laser architectures, we demonstrated extended-cavity Fabry-Perot and mode-locked lasers \cite{kiewiet2026micro}. However, no single-mode or tunable lasers have yet been shown using butt-coupled MTP.

In this work, we demonstrate, for the first time, extended-cavity single-mode lasers using butt-coupled MTP, as outlined in Fig.~\ref{fig:introduction}. We leverage two single-mode extended-cavity architectures engineered to fulfill the diverse requirements of atomic physics: a compact microgear laser utilizing a distributed Bragg reflector microring resonator (DBR-MRR) with a low intrinsic linewidth of 3~kHz and a record-high single-mode output power of $>$22~mW at 794~nm, and widely tunable single-mode Vernier lasers with a 9~nm coarse tuning range around the 795~nm wavelength, $>$140~GHz quasi-continuous fine tuning, and a 45~GHz mode-hop-free tuning range. Together, these lasers demonstrate a complete, highly scalable, highly efficient, and low-noise integration platform capable of replacing macroscopic ECDLs for spectroscopy and state preparation for next-generation Rb quantum processors and atomic clocks.

\section{\label{sec:results} Results}
\subsection{\label{sec:platform} Heterogeneous laser integration using butt-coupled micro-transfer printing}
\begin{figure}[h]
    \centering
    \includegraphics[]{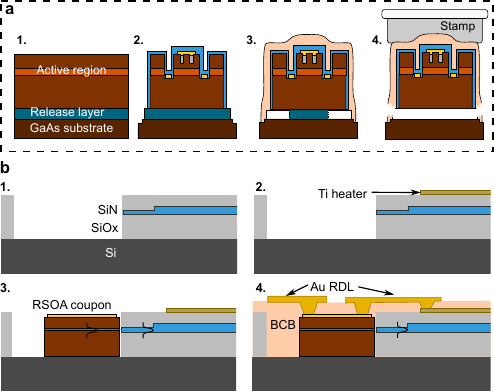}
    \caption{\textbf{III-V coupon preparation and integration process.} \textbf{a.} III-V process for coupon fabrication with: \textbf{1.} epitaxial layer stack; \textbf{2.} laser definitions including waveguide ridge, facet etch and coating, and metal contacts; \textbf{3.} photoresist encapsulation and release etch, leaving free-standing coupons; \textbf{4.} pick-up of the coupon using an elastomer stamp, breaking the photoresist tethers. \textbf{b.} integration process on the SiN wafer with: \textbf{1.} recess etch in the SiN wafer; \textbf{2.} Ti heater definition; \textbf{3.} micro-transfer printing of the III-V coupon in the recess; \textbf{4.} BCB planarization, via etching, and RDL metallization.}
    \label{fig:integration}
\end{figure}

Here, we build further on the heterogeneously integrated (Al)GaAs/SiN photonic platform proposed in previous work \cite{kiewiet2026micro}. The III-V fabrication process is outlined in Fig.~\ref{fig:integration}a. An epitaxial layer stack optimized for butt-coupling with a thick n-contact layer situated on an InGaP release layer is processed into edge-emitting RSOAs with top-side n-contacts. The RSOAs feature a shallow-etched waveguide ridge with etched facets. The rear facet of the RSOAs is coated in a gold highly-reflective (HR) coating and is perpendicular relative to the waveguide propagation direction to maximize reflection. The front facet is coated with a SiN anti-reflective (AR) coating and is angled at 7$^\circ$ to prevent any back-reflection into the laser mode that can destabilize the laser performance. After laser processing, RSOAs are encapsulated in a thick ($>5~\mathrm{\mu m}$) photoresist layer and released from the source substrate by etching the InGaP release layer, leaving free-standing laser coupons tethered to the substrate by the photoresist. 

The laser integration process is shown in Fig.~\ref{fig:integration}b. Ti-based heaters are added to a SiN waveguide platform with recesses etched down to the silicon substrate for butt-coupled MTP integration. The 5~$\mathrm{\mu m}$-thick RSOA ``coupons'' are then picked up using an elastomer stamp, breaking the photoresist tethers, and are micro-transfer printed into the etched recesses. Vertical alignment is ensured by control of the epitaxial layer thicknesses and depth of the recess etch. After RSOA integration, the chips are over-clad and planarized using a benzocyclobutene (BCB) layer. Finally, electrical vias are etched in the BCB to access the laser and heater contacts and a gold redistribution layer (RDL) is added.

The heterogeneously integrated RSOA coupled to a SiN waveguide is a powerful engine to create stable, single-mode lasers. To complete the single-mode lasers, the cavity must be completed in the SiN circuit, providing feedback to the RSOA and wavelength filtering. The strength of the optical feedback, combined with the RSOA gain bandwidth, determines the possible lasing wavelengths. The net internal gain of the RSOA as a function of injection current is shown in Supplementary Fig.~S1. By filtering the optical feedback to only one wavelength within the lasing bandwidth, stable single-mode lasing can be achieved. Furthermore, by increasing the photon lifetime in the cavity using low-loss SiN waveguides/resonators, the optical linewidth of the lasers can be decreased, resulting in more stable and spectrally pure laser output. In this work, this is achieved using two different architectures, which are outlined in the next sections.

\subsection{\label{sec:microgear} Highly efficient and low-noise single-mode microgear laser}

\begin{figure}[h]
    \centering
    \includegraphics[]{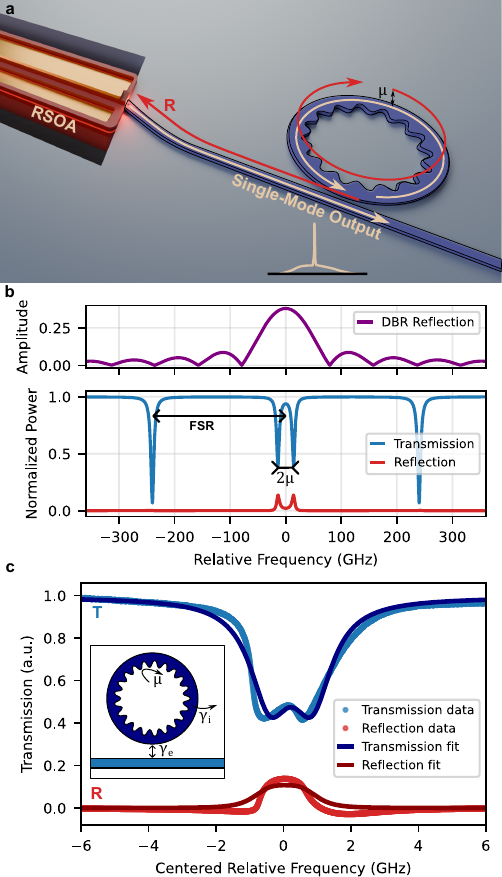}
    \caption{\textbf{Working principle and measurement of the microgear laser.} \textbf{a.} schematic of the laser cavity; \textbf{b.} simulated example reflection and transmission spectra of a microgear reflector showing a single split and reflecting mode; \textbf{c.} measurement of the microgear reflector with inset: model schematic of the reflector.}
    \label{fig:microgear_explanation}
\end{figure}

For the first architecture, we create a narrow-linewidth and highly efficient single-mode laser using a microgear reflector/DBR-MRR \cite{arbabi2011realization}, named after the gear-like corrugated resonator geometry first demonstrated by Fujita and Baba \cite{fujita2002microgear}. This resonant reflector combines the wavelength selectivity of a distributed Bragg reflector (DBR) with the long photon lifetime and compact ($\mathrm{200\ \mu m \times 200\ \mu m}$) footprint of a microring resonator (MRR). It consists of an MRR which has a Bragg grating inscribed in the inner waveguide sidewall, as shown in Fig.~\ref{fig:microgear_explanation}a. The Bragg period is chosen such that an integer number of periods fit in one ring circumference to eliminate phase errors due to discontinuities in the grating perturbation. As the number of periods in one ring circumference is very large ($>$3000), this rounding error does not meaningfully change the free-spectral range (FSR) of the MRR. At the MRR resonances, the amplitude of the field coupled into the resonator will build up, creating a long photon lifetime and a sharp spectral peak. The Bragg grating in the sidewall of the microgear couples the forward and backward modes of the resonator by resonantly reflecting light. If the amplitude of the sidewall corrugation is chosen sufficiently small, the reflection bandwidth can be made smaller than the FSR of the resonator, as illustrated in Fig.~\ref{fig:microgear_explanation}b. This causes only one longitudinal mode of the resonator to experience reflection. The resulting coupling between the forward and backward mode causes a splitting of the resonance into two hybridized modes which both show a strong reflection. As this reflection is highly spectrally selective, it can be used to create a fully single-mode laser. This can be done by using the reflector to self-injection lock a laser \cite{ulanov2024synthetic} to achieve single-mode operation or even to create micro-combs. Alternatively, single-mode lasing can be achieved in a less complex way by creating an extended-cavity laser using the reflector as a frequency-selective cavity mirror \cite{reep2025compact}. The latter strategy is chosen here, as it pairs well with the micro-transfer printed RSOA coupons to create a very compact single-mode laser without adding any fabrication complexity.

\begin{figure*}[ht] 
    \centering
    \includegraphics[]{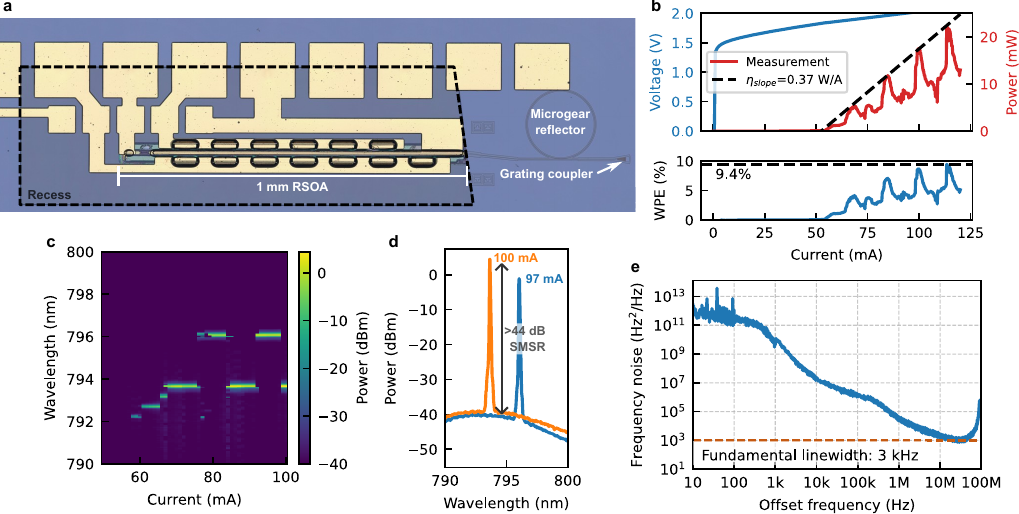}
    \caption{\textbf{Results of the microgear laser} with \textbf{a.} microscope image of the fully-processed laser; \textbf{b.} LIV curve of the microgear laser as well as the corresponding wall-plug efficiency (WPE); \textbf{c.} lasing spectra as a function of the laser driving current; \textbf{d.} lasing spectra of the two dominant laser modes; and \textbf{e.} frequency-noise spectrum of the lasing mode at 100~mA.}
    \label{fig:microgear_laser_results}
\end{figure*}
To test the performance of the microgear reflector, the transmission and reflection of the reflector used were characterized using a Toptica ECDL and a fiber Mach-Zehnder interferometer frequency reference. This is described in more detail in Supplementary Note 2. The reflection and transmission are measured using an on-chip test structure, as also shown in this Supplementary Note. The reflector was designed to have minimal reflection to decrease the bandwidth of the Bragg reflection to minimize the number of resonances that show reflection. To this end, a designed grating amplitude of only 5~nm was chosen. The passive structures were fabricated from pre-annealed 300-nm-thick low-pressure chemical vapor deposition (LPCVD) SiN layers on oxide and patterned using electron-beam lithography (EBL). After etching, a top cladding of SiO\textsubscript{2} was added. The transmission and reflection of the microgear were measured using a reference structure and are shown in Fig.~\ref{fig:microgear_explanation}c. To the measurement data, a temporal coupled-mode theory (t-CMT) model was fitted \cite{li2016backscattering}. This model is illustrated in the inset, where $\mu$ is the reflection rate, $\gamma_{t}$ is the total loss rate including external (coupler) loss rate ($\gamma_e$) and internal loss rate ($\gamma_{in}$). This model is described in more detail in Supplementary Note 3. The minimal amplitude of the Bragg grating is evident from the small splitting of the modes in the transmission spectrum. Additionally, in the reflection spectrum, the two modes cannot be distinguished. Although the overlap of the modes widens the reflection bandwidth slightly, lowering the overall $Q$-factor of the reflection, this has the advantage that there is only one reflection peak, which eliminates any mode-hopping destabilization between the two hybridized reflection modes. The fit results in an internal loss rate of $\mathrm{0.52~GHz}$ and an external loss rate of $\mathrm{0.20~GHz}$, indicating the resonator is in the under-coupling regime. This regime was chosen to maximize the transmission of the reflector to allow for high-output-power operation. The fitted reflection rate is $\mu=\mathrm{0.67~GHz}$ which is very similar to the total loss rate $\gamma_{t}=\gamma_{e}+\gamma_{in}=\mathrm{0.72~GHz}$, explaining the nearly indistinguishable resonances. The microgear reflector has a loaded Q-factor of 261k (as calculated from the fitted loss rates), corresponding to an intrinsic Q-factor of 364k. This quality factor is limited by the sidewall scattering losses in the microgear, as well as the scattering losses introduced by the inscribed sidewall grating, which can be improved by moving from the R\&D process environment to a commercial low-loss platform. 

After SiO\textsubscript{2} top cladding deposition, the RSOA is integrated, resulting in the laser shown in Fig.~\ref{fig:microgear_laser_results}a. The output of the laser is coupled out of the chip using a grating coupler. The laser is very compact, measuring approximately $1.5~\text{mm} \times 0.25~\text{mm}$. The LIV-curve is shown in Fig.~\ref{fig:microgear_laser_results}b along with the WPE curve. As the driving current increases above the threshold of 55~mA, the output power oscillates significantly. This is due to the longitudinal modes of the cavity tuning into and out of the reflecting modes of the microgear reflector. This is because the roundtrip phase of the laser cavity changes as the current is swept. Because the reflector also has a very strong phase response, the wavelength reflected by the microgear can still be in phase with the laser cavity by causing lasing slightly off-resonance. This oscillatory effect could be mostly negated by adding a tunable phase section to the cavity using for example a microheater. If this added phase is chosen optimally for all driving currents, the LI curve would follow the peaks of the present LI curve. These peaks follow a linear slope efficiency of 0.37~W/A, indicating very efficient lasing for a heterogeneous extended-cavity device, with no thermal rollover visible until at least 120~mA. From the LI curve, the coupling efficiency is difficult to estimate, since we cannot be sure when the microgear laser is emitting at the microgear resonance. In previous work, however, this efficiency was extracted at approximately 50~\% \cite{kiewiet2026micro}. The maximum WPE of the laser was measured to be 9.4~\% at 115~mA driving current. This value approaches that of commercial distributed feedback (DFB) lasers at this wavelength.
From the spectral evolution graph in Fig.~\ref{fig:microgear_laser_results}c, after a startup phase until 70~mA, the laser seems to switch between two lasing modes: the designed reflection at 793.6~nm and another mode at 796.1~nm. This means that the microgear also has a significant reflection at a wavelength not fully resonant with the inscribed DBR grating. This is most likely due to a side-lobe of the DBR response, as also seen in previous work \cite{reep2025compact}. From the spectra of the two lasing modes in Fig.~\ref{fig:microgear_laser_results}d, a side-mode suppression ratio (SMSR) of $>$44~dB can be measured, limited by the background of amplified spontaneous emission (ASE) of the RSOA. 
The frequency noise spectrum of the 100~mA lasing mode, measured using an OEwaves OE4000 noise analyzer, is shown in Fig.~\ref{fig:microgear_laser_results}e. From this graph, a narrow fundamental linewidth of 3.0~kHz can be extracted. As this frequency-noise measurement was performed at a fixed driving current, it confirms that at a fixed operating point the laser runs stably on a single longitudinal mode; the output-power oscillations observed in the LI curve therefore reflect mode transitions as the bias is swept rather than an instability at any single operating current. This linewidth is much narrower than typical DFB demonstrations and lower than the linewidths of the Rb resonances around the D1 line, making it suitable for sub-Doppler cooling and low-noise state preparation. This is achieved while keeping a similar efficiency as DFB lasers by leveraging directly butt-coupled RSOA coupons. By adding a phase section heater and a heater to the microgear reflector, more control over the output power and the lasing wavelength can be gained. This would make the laser fully ready for quantum computing applications using the Rb D1 line.
To improve the performance of the microgear laser, one can change the coupling rate between the reflector and the bus waveguide. A larger coupling rate would result in higher feedback into the RSOA, leading to a narrower linewidth at the cost of output power/efficiency. The reverse can also be done by decreasing the reflector-bus-waveguide coupling to increase the slope efficiency of the laser, although this will also increase the linewidth and threshold current. By using a more optimized SiN platform, quality factors up to 2 million can be reached \cite{ulanov2024synthetic}. As the intrinsic or Schawlow-Townes linewidth scales as $\Delta\nu_{\mathrm{ST}}\propto1/Q^2$ \cite{fan2020hybrid,coldren2012diode}, this would result in a linewidth on the order of 50~Hz, making the laser suitable for high-precision metrology.
With higher quality factors, microgear lasers can possibly be used to generate solitons at similar on-chip powers as generated in this work \cite{yu2021spontaneous} to make compact, efficient frequency comb sources. This requires pump powers up to 20~mW, which is attainable for the laser shown in this work. This would require precise engineering of dispersion in combination with the back-scattering rate, which is controlled by the DBR corrugation amplitude. The advantages and viability of using an RSOA instead of a DFB, as used in previous demonstrations \cite{ulanov2024synthetic}, are still to be determined.

\subsection{\label{sec:vernier} Widely-tunable single-mode lasers}
\begin{figure}[h]
    \centering
    \includegraphics[]{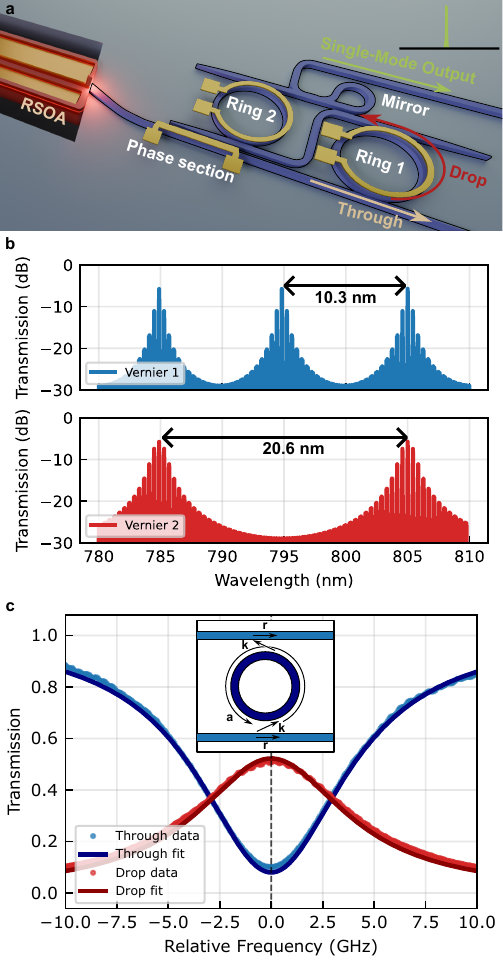}
    \caption{\textbf{Explanation of the Vernier laser} with \textbf{a.} schematic of the Vernier laser cavity including microheaters; \textbf{b.} modeled transmission spectra of the two Vernier filters used in this work; and \textbf{c.} transmission measurements of the microring resonator with inset: transfer-matrix model explanation used to fit to the measurement data.}
    \label{fig:vernier_explanation}
\end{figure}
\begin{figure*}[ht]
    \centering
    \includegraphics[]{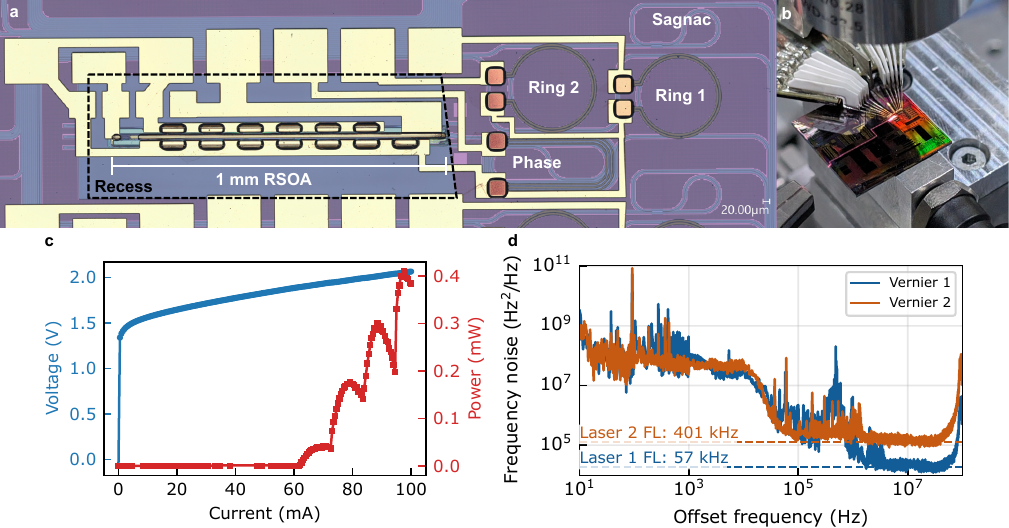}
    \caption{\textbf{Vernier laser results} showing \textbf{a.} microscope picture of a fully-processed Vernier laser; \textbf{b.} camera picture of a Vernier laser operating showing scattering of light on the laser and the routing to the chip edge; \textbf{c.} LIV curve of Vernier laser 1 with constant heater powers; and \textbf{d.} frequency noise spectra of both Vernier lasers along with their fundamental linewidth.}
    \label{fig:vernier_results}
\end{figure*}

A more common and flexible option for creating a single-mode extended-cavity laser is by using a Vernier filter in the extended feedback cavity. This filter can be combined with a broad-band Sagnac loop mirror to create the extended cavity of the laser, as shown in Fig.~\ref{fig:vernier_explanation}a. This has previously been shown at wavelengths all over the spectrum~\cite{fan2020hybrid, winkler2024widely,franken2025widely} using hybrid integration, where a III-V RSOA chip is butt-coupled to a low-loss feedback chip with a Vernier filter. Demonstrations using more scalable, fully heterogeneously integrated III-V semiconductor optical amplifiers (SOAs) evanescently coupled via adiabatic tapers are mostly limited to telecommunication (1550~nm) and datacom (1310~nm) wavelengths \cite{hulme2013widely, pan2024iii, malik2020widely}, with some demonstrations at shorter wavelengths as described earlier. However, no previous demonstration has been shown using butt-coupled MTP integration.

By cascading two MRRs in add-drop configuration with slightly different radii, the Vernier effect is used to make a filter with an FSR that can far exceed those of the individual rings. To have maximum transmission for the cascaded rings, both rings must be resonant at the same wavelength. Since the FSR of both rings differs slightly, the resonances adjacent to the shared resonant wavelength do not fully overlap anymore, as shown in Fig.~\ref{fig:vernier_explanation}b. The FSR of the Vernier filter, $FSR_{\mathrm{Vernier}}$ is then determined by the difference of the FSRs of the individual rings (rings 1 and 2) as
\begin{equation}
    FSR_{\mathrm{Vernier}} = \frac{FSR_1\times FSR_2}{|FSR_1-FSR_2|}.
\end{equation}

Two different lasers were designed for this work, where the main difference is the Vernier FSR: 10.3~nm and 20.6~nm. These values were chosen to best match the gain bandwidth of the RSOA. The SiN used for the Vernier lasers was fabricated on a 200-mm SiN platform of imec developed for the VISSION platform, featuring waveguide losses of $<$0.6~dB/cm at the 800~nm wavelength. The MRRs were characterized using test structures. The results of this characterization are shown in Fig.~\ref{fig:vernier_explanation}c. The drop and through responses are fitted using a shared cost function using the model shown in the inset, where $r$ and $k$ are the ring-bus coupler transmission and coupling ratios, and $a$ is the microring roundtrip transmission coefficient \cite{bogaerts2012silicon}. This model is described in more detail in Supplementary Note 4. Unfortunately, the fabricated ring resonators exhibited higher-than-anticipated propagation losses. This resulted in a sub-optimal loaded quality factor of approximately 40k, reducing laser noise performance. Furthermore, this excess ring loss also limits the drop-port coupling efficiency to a little over 50~\%. Since this loss is incurred four times in each cavity roundtrip (double pass through both rings), this adds a loss of almost 12~dB to the cavity, significantly limiting the performance of the lasers. 

\begin{figure*}[ht]
    \centering
    \includegraphics[]{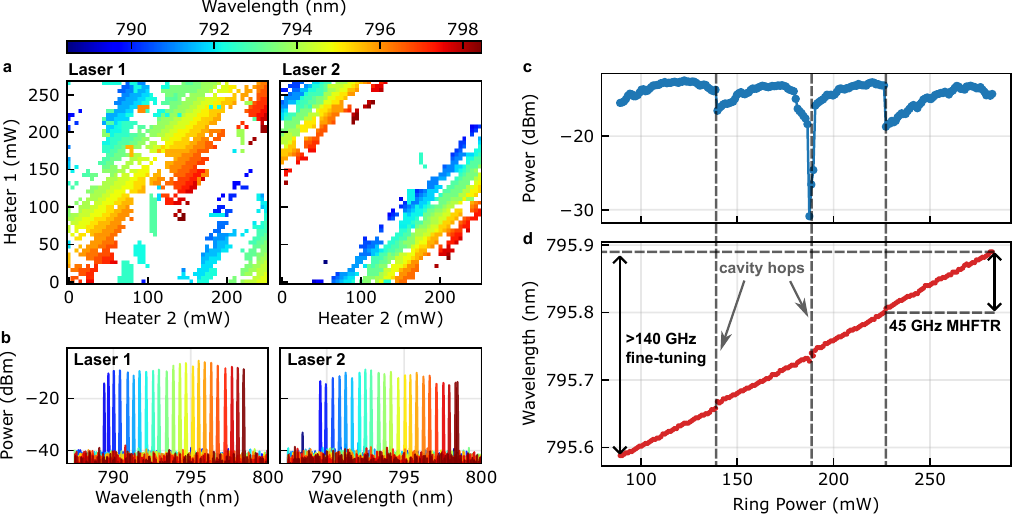}
    \caption{\textbf{Vernier laser tuning characterization.} \textbf{a.} coarse tuning wavelength maps for Vernier laser 1 and Vernier laser 2, showing the lasing wavelength as a function of the two heater powers. Operating points without lasing are colored white; \textbf{b.} example overlaid coarse tuning spectra for Vernier laser 1 and Vernier laser 2. Fine tuning for a simultaneous sweep of the ring heater powers for Vernier laser 1 showing \textbf{c.} peak power and \textbf{d.} lasing wavelength as function of the power supplied to ring 1. The grey dashed lines indicate the laser hopping between adjacent longitudinal cavity modes.}
    \label{fig:vernier_coarse_tuning}
\end{figure*}

The fully-processed Vernier lasers are shown in Fig.~\ref{fig:vernier_results}a, featuring the 1-mm-long RSOA in the etched recess, the two microrings forming the Vernier filter along with their microheaters, a phase section with a microheater, and a Sagnac loop mirror with 75~\% reflection. The output of the laser is routed to the edge of the chip---along with all the monitor ports---and terminated using an angled facet edge coupler to reduce parasitic reflections into the laser cavity. The light of the laser is then extracted using a lensed fiber, as shown in Fig.~\ref{fig:vernier_results}b, and an edge coupler loss of 6.0~dB is measured using a reference waveguide. Two Vernier lasers are characterized in this work: Vernier laser 1 with a Vernier FSR of 10.3~nm and Vernier laser 2 with a Vernier FSR of 20.6~nm. The LIV curve of Vernier laser 1 is shown in Fig.~\ref{fig:vernier_results}c, showing a maximum output power of 0.4~mW. The stark discrepancy with the output power of the microgear laser is due to the high losses in the Vernier cavity attributed to the MRRs. Despite these losses, both lasers show single-mode lasing. Their frequency noise spectra are shown in Fig.~\ref{fig:vernier_results}d. Vernier laser 1 shows the best fundamental linewidth of 57~kHz, whereas Vernier laser 2 shows a higher fundamental linewidth of 401~kHz. This difference is most likely due to a variation in coupling efficiency of the RSOA, which is also why Vernier laser 2 shows a lower output power.

The difference between the two lasers is clearest in the Vernier tuning maps in Fig.~\ref{fig:vernier_coarse_tuning}a. Here, the power to the two microring heaters is swept in a 2D grid. The lasing wavelength is plotted using a colormap, where a white data point means there was no lasing at this working point. Both lasers show a similar tuning range of around 9~nm, which is mostly limited by the high losses in the laser cavity. The lasing map shows holes in areas where lasing should be expected. This is due to the short length of the laser cavity causing an additional Vernier effect between the MRR Vernier filter and the laser cavity. As the heater powers in the ring change, the ring resonances shift, but also a phase shift is induced due to thermal cross-talk. Only if the total phase of the cavity is a multiple of $2\pi$, or close to it, lasing is possible. By changing the power to the phase section heater, this phase shift can be optimized and the holes in the lasing map can be filled in. The full coarse tuning range of both lasers is displayed in Fig.~\ref{fig:vernier_coarse_tuning}b.
If diagonal lines in the lasing map are followed, meaning the power to both heaters is changed equally, the ring resonances shift in tandem. This can be used to fine-tune the laser wavelength. In order to reach the largest fine-tuning range, the cavity phase should also be changed accordingly to always achieve a resonant reflection to the RSOA. To do this, the optimal ratio of change in the phase heater power to the change in the ring heater powers was found to be
\begin{equation}
    \frac{\Delta P_{\mathrm{phase}}}{\Delta P_{\mathrm{ring}}}= 2.5.
\end{equation}
The resulting fine-tuning is shown in Fig.~\ref{fig:vernier_coarse_tuning}c and d. A fine-tuning range exceeding 0.3~nm or 140~GHz was shown, limited by the cavity phase heater efficiency. This could in future be mitigated by extending the phase heater section to spread out the dissipated power in the heater, or by under-cutting of the waveguide under the heater. Within this tuning range, a number of mode hops can be seen, most clearly from the maximum power. As the wavelength changes minimally during such a hop, the laser is hopping between adjacent cavity modes. The wavelength is still closely tracking the synchronized tuning ring resonators, so this extended range with small mode hops is still usable for certain spectroscopic applications. A maximum fully mode-hop-free tuning range (MHFTR) of 45~GHz is visible, likely limited by the cross-talk between the ring and phase heaters, as well as the wide ring resonances, but is comparable with benchtop ECDLs. 
The results of the Vernier laser clearly showcase the possibility to create complex extended-cavity laser designs with the directly butt-coupled MTP laser integration approach. While current output powers and linewidths are limited by the sub-optimal SiN cavity design, the laser is already suitable for spectroscopic applications and locking to Rb atoms to provide an absolute frequency reference, as the linewidth is sufficiently narrower than those of the relevant Rb lines \cite{heavens1961radiative}. This means the laser can be used to probe the Rb to monitor and find resonances, as well as to have an absolute frequency reference by locking it to the Rb D1 line. The SiN cavity design can easily be improved in future iterations to reach wider tuning ranges up to 20~nm, ultimately limited by the gain bandwidth of the RSOA. Furthermore, optimizing the MRR drop port transmission from 3~dB down to 0.5~dB would decrease the intra-cavity roundtrip loss by 10~dB. According to the analysis in \cite{kiewiet2026micro}, this will decrease the threshold down to about 50~mA and unlock 10-mW-level output powers. By decreasing the Sagnac mirror reflectivity from 75~\% down to 50~\%, the output power could be increased to 20~mW, at the cost of a slightly higher fundamental linewidth. Optimizing the ring quality factor to 1 million, applying the same Schawlow-Townes scaling as for the microgear laser, would decrease the intrinsic linewidth from 57~kHz down to 90~Hz, unlocking high-precision time-keeping applications.

\section{Discussion}

\begin{table*}[t]
\centering
\caption{\textbf{Heterogeneous-integration laser comparison in the 780--800~nm Rb band.} Values marked ``—'' are not reported in the cited work.  $^\dagger$Reduced to 92.4~Hz with self-injection locking. $^\ddagger$Off-chip power; on-chip value not reported. WB\,=\,wafer bonding; MTP\,=\,micro-transfer printing; ECFP\,=\,Extended-Cavity Fabry-Perot.}
\label{tab:comparison}
\begin{tabular}{lllcccccc}
\toprule
Work & Laser type & Integration & \makecell{$\lambda$\\(nm)} & \makecell{On-chip power\\(mW)} & \makecell{WPE\\(\%)} & \makecell{Fund. linewidth\\(kHz)} & \makecell{Coarse tuning\\(nm)} & \makecell{MHFTR\\(GHz)} \\
\midrule
This work (Microgear) & DBR-MRR & MTP & 794 & $>$22 & 9.4 & 3.0 & fixed & — \\
This work (Vernier)   & Vernier & MTP & 795 & 0.4 & 0.2  & 57 & 9 & 45 \\
Kiewiet 2026 \cite{kiewiet2026micro} & ECFP & MTP & 796 & 4.4 & 2.2 & — & — & — \\
\midrule
Zhang 2023 \cite{zhang2023photonic} & FP & WB & 780 & 12 & 6 & — & — & — \\
Zhang 2023 \cite{zhang2023photonic} & Vernier & WB & 780 & — & — & 4.35$^\dagger$ & $\sim$18 & — \\
Castro 2025 \cite{castro2025integrated} & Vernier & WB & 765--795 & $>$10 & — & 5.85 & 20 & 108 \\
Thiel 2026 \cite{thiel2026wafer} & Vernier & WB & 780 & $>$10$^\ddagger$ & — & ``kHz-level'' & 25 & — \\
\bottomrule
\end{tabular}
\end{table*}

In summary, we have demonstrated a highly versatile, wafer-scale compatible toolkit of heterogeneously integrated lasers at 795~nm tailored specifically for Rb-based quantum technologies. By leveraging MTP, we directly butt-coupled pre-fabricated GaAs-based RSOA coupons into recesses on a SiN waveguide platform. This approach enables highly efficient optical coupling while leveraging superior thermal dissipation directly to the silicon substrate to enable high-power operation. This is clearly apparent from the compact microgear laser which exhibited high power and highly efficient single-mode lasing exceeding 22~mW at a 9.4~\% WPE, with an intrinsic linewidth of 3~kHz. This performance exceeds any previous heterogeneously integrated demonstration at this wavelength in on-chip output power and wall-plug efficiency, as multi-wavelength Fabry-Perot waveguide-coupled lasing powers have been reported up to 12~mW at 6~\% WPE \cite{zhang2023photonic} and single-mode lasing up to 10~mW with no reported WPE \cite{thiel2026wafer}, as summarized in Table~\ref{tab:comparison}. Furthermore, we show the possibility to create more complex Vernier filter-based tunable lasers using this integration method, which achieved a coarse tuning range of 9~nm and fine-tuning exceeding 140~GHz. Although output power and noise performance are currently limited by the SiN MRR performance, even in their current iteration, these lasers are well-suited for spectroscopic applications and direct frequency locking to the Rb D1 line. 

Looking forward, optimizing the SiN cavity design presents clear pathways to significantly enhance laser performance. For the microgear architecture, incorporating a dedicated phase section heater and a microgear heater will yield precise control over both output power and lasing wavelength. This would enable a more continuous LI-curve with a much larger mode-hop-free current operation range, as well as fine-tuning of the lasing wavelength to allow interaction with narrow-linewidth atomic resonances. The as-fabricated microgear modes at 793.6 and 796.1~nm lie within approximately 1.4~nm of the Rb D1 line. Since the absolute wavelength is set lithographically, closing this gap is primarily a matter of re-targeting the grating period in a subsequent mask revision; a microgear heater with sufficient tuning efficiency could then bridge the residual fabrication offset and hold the laser on the atomic resonance. Furthermore, utilizing an ultra-low-loss SiN platform could unlock quality factors up to 2 million. This enhancement could drastically reduce the fundamental linewidth to the 50~Hz regime and, with some dispersion control, enable on-chip soliton generation for compact frequency combs. For the Vernier lasers, reducing excess intra-cavity losses will unlock output powers at the 10-mW level, while improving the ring quality factors can yield sub-kHz-level fundamental linewidths. Additionally, by optimizing the feedback cavity loss, the tuning range could be expanded up to 20~nm to fully exploit the RSOA gain bandwidth. Ultimately, this integration strategy offers a highly scalable solution to replace macroscopic ECDLs, bringing cost-effective fully integrated, field-deployable optical control systems for sub-Doppler cooling, high-fidelity state preparation, spectroscopy, and optical atomic clocks within reach.

\section{Materials and Methods}
\subsection{III-V RSOA coupon fabrication}
RSOA coupons are fabricated on dies taken from 2-inch GaAs wafers with epitaxial layers grown using metal-organic vapor phase epitaxy. The epitaxial layer stack is detailed in \cite{kiewiet2026micro}. All (Al)GaAs etching is done using BCl\textsubscript{3}/H\textsubscript{2}-based plasma in an inductively-coupled plasma (ICP) etcher. For this etching, a neutral-stress SiN hard-mask is used, deposited using plasma-enhanced chemical vapor deposition and patterned using UV lithography and SF\textsubscript{6}/CF\textsubscript{4}/H\textsubscript{2}-based reactive-ion etching (RIE). This process is used to pattern the shallow-etched waveguide, vias to access the n-doped contact layer, and the laser facets. All etched surfaces are passivated using SiN following a wet native oxide removal using dilute HCl. The contacts---Ti/Au for the p-contact and Ni/Ge/Au for the n-contact---are deposited using electron-beam evaporation following another native oxide removal using dilute HCl. The metal contacts are patterned using a lift-off process and annealed using rapid-thermal annealing at 430~$^{\circ}$C. A gold mirror on a $\sim$1-nm-thick Ti adhesion layer is added to the passivated rear facet using angled electron-beam evaporation and lift-off to function as an HR coating. After the laser coupon is finished, the InGaP release layer is patterned using a cyclical (to prevent over-heating) BCl\textsubscript{3}/H\textsubscript{2}-based ICP etch and a resist mask. Subsequently, after a dry plasma-based native oxide removal of the substrate, photoresist encapsulation is deposited and patterned using UV lithography to form tethers. Lastly, a 1~M HCl solution is used to selectively under-etch the laser coupons by removing the InGaP release layer. This is followed by extensive rinsing in a cross-flow bath to eliminate any HCl residue that will cause local oxidation.

\subsection{Laser Integration using Micro-Transfer Printing}
The MTP process is described in more detail in \cite{kiewiet2026micro}. The GaAs RSOA coupons are integrated on the imec SiN platform to create tunable single-mode lasers and on the in-house EBL platform to create the high-power single-mode microgear laser. On the imec platform, recesses for butt-coupled MTP integration are already included in the silicon photonics process. To create the microgear lasers, a pre-annealed 300-nm-thick SiN layer on top of thermal oxide is patterned using EBL and RIE etching. A top cladding of SiO\textsubscript{2} is added using ICP-chemical vapor deposition. Subsequently, recesses are etched through the SiO\textsubscript{2} and SiN layers using a highly physical CHF\textsubscript{3}/Ar-based etch. On the imec platform, 250-nm-thick Ti heaters are patterned using UV lithography and a lift-off process. A thin Au cap is added to the heaters to protect the Ti from oxidation. To allow better MTP yield, a 50-nm-thick photosensitive BCB adhesion layer is patterned in the recess. The patterning of this adhesion layer allows for the removal of the ramp of the BCB at the edge of the recess. After MTP, the entire sample is covered in thick BCB, which is etched back to approximately 1~$\mathrm{\mu m}$ above the laser coupon. This planarizes the chip and fills any voids in the optical path. Lastly, vias are etched in the BCB and a gold RDL is added to allow for easy electrical probing of the laser gain and the heaters.

\subsection{Laser characterization}
The passives (microgears and MRRs) are characterized using grating couplers and an ECDL (Toptica DL pro) with a fiber Mach-Zehnder interferometer frequency reference. The lasers are characterized on a temperature-controlled chuck, where they are held down using vacuum. No sub-mount or thermal paste is used to aid heat sinking. The microgear laser output is coupled into a cleaved HP780 single-mode fiber using a grating coupler. The transmission spectrum of the grating coupler is measured using a short reference waveguide. To determine the LI curve, the fiber-coupled light is measured using a (Thorlabs S120C) Si power sensor. Furthermore, the spectrum is measured at intervals of 1~mA driving current using an (Anritsu MS9740A) optical spectrum analyzer (OSA) using a resolution of 0.03~nm. The waveguide-coupled power is determined by subtracting the grating coupler loss at the lasing wavelength for each driving current measured this way. The light of the Vernier laser is captured using a lensed fiber and an angle edge-coupler to reduce reflections. A reference waveguide is used to measure the edge coupler loss. The Vernier tuning maps are extracted using the same OSA with a resolution of 0.03~nm. The smaller frequency shifts are measured using over-sampling of the OSA spectrum. The frequency noise of all lasers is measured using a laser linewidth analyzer (OEwaves OE4000 HI-Q).

\section*{Data availability}
The data that support the findings of this study are available from the corresponding author upon reasonable request.

\bibliography{bibliography}
\begin{acknowledgments}
We acknowledge funding by the Horizon Europe programme of the European Union (Grant agreement ID: 101070622) and the Flemish Research Council (FWO PhD fellowship grant 1SF9322N).
\end{acknowledgments}

\section*{Author contributions}
M.K., S.C., and B.K. conceived the idea of the project. M.K. and S.C. designed the III-V coupons. M.K. and S.C. fabricated the III-V coupons with assistance from K.A. M.K. designed and simulated the SiN components and circuits and developed the integration technique. T.R. contributed to the microgear reflector design. M.K. measured and characterized the lasers with assistance from K.V.G. and B.K. G.J., M.R., P.V.D., and R.J. developed imec's 200-mm SiN photonics platform and provided the wafer. M.K. prepared figures and wrote the manuscript. All authors reviewed the manuscript. G.R., K.V.G., and B.K. supervised the project.

\section*{Competing interest statement}
The authors declare no competing interests.

\end{document}